\documentclass[english]{article}
\usepackage[LGR,T1]{fontenc}
\usepackage[latin9]{inputenc}
\usepackage{geometry}
\usepackage{textcomp}
\usepackage{amsmath}
\usepackage{amssymb}
\usepackage{setspace}

\makeatletter

\DeclareRobustCommand{\greektext}{%
  \fontencoding{LGR}\selectfont\def\encodingdefault{LGR}}
\DeclareRobustCommand{\textgreek}[1]{\leavevmode{\greektext #1}}
\DeclareFontEncoding{LGR}{}{}
\DeclareTextSymbol{\~}{LGR}{126}
\providecommand{\tabularnewline}{\\}

\makeatother

\usepackage{babel}
\begin{document}

\title{\textbf{Testing for Characteristics of Attribute linked}\\
\textbf{Infinite Networks based on Small Samples }}

\author{\textbf{Koushiki Sarkar and Diganta Mukherjee}\\
\textbf{Indian Statistical Institute, Kolkata}}

\maketitle
\medskip{}

\begin{abstract}
The objective of this paper is to study the characteristics (geometric
and otherwise) of very large attribute based undirected networks.
Real-world networks are often very large and fast evolving. Their
analysis and understanding present a great challenge. An Attribute
based network is a graph in which the edges depend on certain properties
of the vertices on which they are incident. In context of a social
network, the existence of links between two individuals may depend
on certain attributes of the two of them. We use the Lovasz type sampling
strategy of observing a certain random process on a graph \textquotedblright locally\textquotedblright ,
i.e., in the neighborhood of a node, and deriving information about
\textquotedblright global\textquotedblright{} properties of the graph.
The corresponding adjacency matrix is our primary object of interest.
We study the efficiency of recently proposed sampling strategies,
modified to our set up, to estimate the degree distribution, centrality
measures, planarity etc. The limiting distributions are derived using
recently developed probabilistic techniques for random matrices and
hence we devise relevant test statistics and confidence intervals
for different parameters / hypotheses of interest. We hope that our
work will be useful for social and computer scientists for designing
sampling strategies and computational algorithms appropriate to their
respective domains of inquiry. Extensive simulations studies are done
to empirically verify the probabilistic statements made in the paper.
\end{abstract}

\section{Introduction}

\subsection{Need for Sampling and Common Strategies }

{\large{}Real-world networks are often very large and fast evolving.
Their analysis and understanding present a great challenge. In the
past few years, a number of different techniques have been proposed
for sampling large networks to allow for their faster and more efficient
analysis. Several studies on network sampling analyze the match between
the original networks and their sampled variants {[}1, 2, 3{]}, as
well as comparing the performance of different sampling techniques
{[}4, 5, 6{]}.}{\large \par}

\medskip{}

{\large{}Sampling techniques can be roughly divided into two categories:
random selection and network exploration techniques. In the first
category, nodes or links are included in the sample uniformly at random
or proportional to some particular characteristic like degree. In
the second category, the sample is constructed by retrieving a neighborhood
of a randomly selected seed node using different strategies like breadth-first
search, random walk and forest-fire. On these basis, the following
algorithms have been proposed in the literature.}{\large \par}
\begin{itemize}
\item {\large{}random node selection {[}4{]} (RNS), where the sample consists
of nodes selected uniformly at random and all their mutual links }{\large \par}
\item {\large{}random node selection by degree {[}4{]} (RND), the nodes
are selected randomly with probability proportional to their degrees
and all their mutual links are included in the sample}{\large \par}
\item {\large{}random link selection {[}4{]} (RLS), where the sample consists
of links selected uniformly at random}{\large \par}
\item {\large{}random link selection with subgraph induction {[}7{]} (RLI),
the sample consists of links selected uniformly at random and any
additional links between their endpoints}{\large \par}
\item {\large{}random walk sampling {[}4{]} (RWS), where the random walk
is simulated on the network, starting at a randomly selected seed
node}{\large \par}
\item {\large{}forest-fire sampling {[}4{]} (FFS). Here, a broad neighborhood
of a randomly selected seed node is retrieved from partial breadth-first
search}{\large \par}
\item {\large{}random walk sampling with subgraph induction (RWI) and }{\large \par}
\item {\large{}forest-fire sampling with subgraph induction (FFI)}{\large \par}
\end{itemize}
\medskip{}

{\large{}Recall the }\textit{\large{}Lovasz}{\large{} sampling strategy
of observing a certain random process on a graph \textquotedblright locally\textquotedblright ,
i.e., in the neighborhood of a node, and deriving information about
\textquotedblright global\textquotedblright{} properties of the graph.
For example, what do you know about a graph based on observing the
returns of a random walk to a given node? Almost all sampling strategies
uses this philosophy or some close variant. We also aim to use this
as an ingredient in our sampling strategy and derive a test for such
returns. }{\large \par}

\subsection{Why Attribute based Network?}

{\large{}An Attribute based network is a graph in which the edges
depend on certain properties of the vertices on which they are incident.
In context of a social network, the existence of links between two
individuals may depend on certain attributes of the two of them, for
example their geographic location or socioeconomic status. We work
with the underlying assumption that similar people connect to each
other with higher probability. In the context of a social or a neural
network, the connection between individual vertices depend on certain
intrinsic qualities of the vertices themselves. It makes sense to
consider the connection probabilities as a function of the vertex
attributes. In earlier work (Sarkar, Ray and Mukherjee, 2015) we have
shown that in the context of predictive modeling, attribute based
networks are indeed worthwhile to study.}{\large \par}

\subsection{Plan of the Paper}

{\large{}The objective of this paper is to study the characteristics
(geometric and otherwise) of very large attribute based undirected
networks. The corresponding adjacency matrix is our primary object
of interest. We study the efficiency of recently proposed sampling
strategies, modified to our set up, to estimate the degree distribution,
centrality measures, planarity etc. The limiting distributions are
derived using recently developed probabilistic techniques for random
matrices and hence we devise relevant test statistics and confidence
intervals for different parameters / hypotheses of interest. We hope
that our work will be useful for social and computer scientists} {\large{}for
designing sampling strategies and computational algorithms appropriate
to their respective domains of inquiry. Extensive simulations studies
are done to empirically verify the probabilistic statements made in
the paper. }{\large \par}

\section{Preliminaries}

{\large{}Let the network be represented by a simple undirected graph
G = (V; E), where V denotes the set of nodes (n = |V|) and E is the
set of links (m = |E|). The goal of network sampling is to create
a sampled network G' = (V'; E'), where V' $\subset$V , E' $\subset$
E and n' = |V| <\textcompwordmark{}< n, m' = |E'| <\textcompwordmark{}<
m. The sample G' is obtained in two steps. In the first step, nodes
or links are sampled using a particular strategy like random selection
and network exploration sampling. In the second step, the sampled
nodes and links are retrieved from the original network.}{\large \par}

\subsection{\emph{Notions of Centrality}}

{\large{}A network can be characterized by various notions of centrality,
whose relevance and utility are context-specific. A complex network
with a heterogeneous topology might not have the same optimality properties
for a single measure of centrality throughout the graph. }{\large \par}

{\large{}However, for a sampling based approach without any prior
idea of the graph type, it may be difficult to know which centrality
measure is best suited for the study of the graph. If we operate under
the simplified assumptions about the attributes-based network then
our graph structure is simplied. Particularly, if the attribute random
variable $\left\{ X_{i}\right\} $take values in a finite set, then
the set of possible connecting probabilities is also finite. Then
we can break a graph into different classes which are expected to
have similar behavior.}{\large \par}

{\large{}Here we intend to develop a sampling analogue for finding
the Eigenvector centrality, which is the solution of the $Eigenvector\, Equation$
to $Ax=\lambda x$. According to the Perron Frobenius theorem due
to strict positivity of the eigen values we only require the largest
eigenvalue.}{\large \par}

{\large{}A possible generalization of Eigenvector centrality as well
as Degree Centrality is the Katz centrality. It measures the number
of all nodes that can be connected via a path to the vertex in question,
while the contributions to distant nodes are devalued. It is mathematically
written as $x_{i}=\sum_{k}\sum_{j}\alpha^{k}(A)_{ji}^{k}$}{\large \par}

{\large{}Katz centrality can be viewed as a variant of eigenvector
centrality. Another form of Katz centrality is $x_{i}=\alpha\sum_{j=1}^{N}a_{ij}(x_{j}+1)$.
Compared to the expression of eigenvector centrality, $x_{j}$ is
replaced by $x_{j}+1$.}{\large \par}

{\large{}It is shown that the principal eigenvector (associated with
the largest eigenvalue of A, the adjacency matrix) is the limit of
Katz centrality as$\alpha$ approaches $1/\lambda$ from below.}{\large \par}

\subsection{Assumptions on the Network}

{\large{}Basic Assumptions: We denote the variable $X_{i}$ as an
attribute of the class $i$, where $i$ is assumed to take only finitely
many values. In a population, there can be infinitely many people
with same attribute. Let us consider all of them members of the class
$i$. Call it $c_{i}$.}{\large \par}
\begin{itemize}
\item {\large{}Looking at degree distribution is not very meaningful as
even if we know that the degrees are distributed by power law or Normally,
we still don't know what the degree should be for an individual node.}{\large \par}
\item {\large{}The degree needs to be a specific property of a node for
us to meaningfully select a node. In context of social networks, it
makes sense to consider vertices appended with attributes. An edge
or connection can be considered to be dependent on the attributes
of the involved nodes.}{\large \par}
\item {\large{}We consider the accessory variable $X_{i}$ appended to each
vertex i.}{\large \par}
\item {\large{}Call the indicator function $\delta_{ij}=\begin{cases}
1ifconnected & 0ow\end{cases}$}{\large \par}
\item {\large{}We need to look at $p_{ij}=P(\delta_{ij}=1|X_{i},X_{j})$.
If Erdos Renyi Graph, then this is the unconditional probability same
for all i,j. }{\large \par}
\item {\large{}We make some assumptions on $p_{ij}.$Even if $i$ and $j$
belong to the same class and hence share the same attributes, $p_{ij}\neq1$.
$p_{ij}$is assumed to be bounded away from 1 and 0.}{\large \par}
\item {\large{}Fix an i. Consider $arg\underset{i}{max}\sum_{j}P(\delta_{ij}=1|X_{i},X_{j})$.
This can be approximated by $arg\underset{i}{max}E(d|X_{i})$ where
d is the degree}{\large \par}
\item {\large{}The $\left\{ X_{n}\right\} $sequence, if stochastic, is
assumed to form a Markov Random Field. we have a sort of dependence
structure within a neighborhood; distant nodes can be assumed to be
more or less independent.}{\large \par}
\end{itemize}
{\large{}In particular, if $\left\{ X_{n}\right\} $ is finite, then
$p_{ij}$also takes finitely many values.}{\large \par}

\section{Probabilistic Formulation}

{\large{}Instead of the Adjacency Matrix, we can consider the matrix
$P=((p_{ij}))$, the matrix of the probabilities. If we consider a
random motion on the graph, we consider $p_{ij}$to be a transition
probability on the graph. A possible notion of centrality in this
context is whether the vertex is recurrent or transient. A recurrent
vertex indicates that there are multiple paths leading back to the
graph. This is somewhat analogous to Betweenness centrality. We need
to however note that in a large graph modeled on a social network
which would be mostly sparse with intermediate densely connected cliques,
the actual betweenness for all but few vertices would be vanishingly
small. On a global scale these vertices may not be important; but
their local influence cannot be dismissed.}{\large \par}

{\large{}We look at an irreducible aperiodic subset of the graph.
If the motion is considered to be Markov, then noting that transience
and recurrence are solidarity properties, we attempt to verify that
using our model.}{\large \par}

{\large{}If $d$ is a metric defined on the $\sigma-Field$ generated
by the random variables $\left\{ X_{n}\right\} $, then consider $p_{ij}=f(d(X_{i},X_{j}))$,
where $f$ is a decreasing function of $d$ and bounded in $[0,1].$An
easy example is $d'(x,y)=1-min(1,d(x,y))$}{\large \par}

{\large{}Again, if $\left\{ X_{n}\right\} $is finite (or countable),
$d$ only takes finitely (countably) many values and consequently
the set of values of $p$ is also finite (countable.)}{\large \par}

{\large{}If the $\left\{ X_{n}\right\} $is known, then $P$ is also
completely known, as is $P^{k}.$ In principal, we can also calculate
if $\sum f_{ii}^{n}<1$, where $f_{ii}^{(n)}=\Pr(T_{i}=n)$. If this
holds, then the vertex is transient, else recurrent.}{\large \par}

{\large{}\bigskip{}
}{\large \par}

\subsection{Degree Distribution}

{\large{}We are estimating $p_{ij}=P(\delta_{ij}=1|X_{i},X_{j}).$
Let $\mathbb{P}=((p_{ij}))$ and $\mathbb{E}=((\delta_{ij}))$, symmetric
with $\delta_{ii}=0,\;\forall i$. So $\delta_{ij}\sim Bernouli(p_{ij})$.
Symbolically $\mathbb{E}\sim Bernouli(\mathbb{P}).$}{\large \par}

{\large{}Let degree, $d_{i}=\sum_{j}\delta_{ij}$. What is the distribution
of $\delta_{ij}|d_{i}$, should be \textquotedbl{}hypergeometric\textquotedbl{}
type?}{\large \par}

{\large{}So can we use }\emph{\large{}iterated expectation}{\large{}
in the following way? }\\
{\large{} First condition on row total $d_{i}$ to use the \textquotedbl{}hypergeometric
type\textquotedbl{} calculation for $(\delta_{ij}|X_{i},X_{j})$.
}\\
{\large{} Then take expectation on $d_{i}\sim F(.|X_{i})$ following
from the }\emph{\large{}ERG}{\large{} model. If we can show that the
distribution of $d_{i}$ is of \textquotedbl{}Binomial type\textquotedbl{},
then for fixed $|V|=n$ we can do the calculation and then take limit
$n\rightarrow\infty$?}{\large \par}

{\large{}Assuming $\{X_{i}\}$to be non-stochastic, if degree of the
$i^{th}$ vertex is $d_{i},$we have:}{\large \par}

{\large{}$d_{i}=\sum_{j=1,i\neq j}^{n}\delta_{ij},$ where $\delta_{ij}$
is the indicator variable which is 1 if there is a connection between
the $i^{th}$ and the $j^{th}$vertices.}{\large \par}

{\large{}If $p_{ij}$is the connection probability of the $i$ and
the $j^{th}$vertex, then $p_{ij}=f(x_{i},x_{j})$ is completely known.}{\large \par}

{\large{}Consider the degree proportion,}{\large \par}

{\large{}$\frac{d_{i}}{n-1}=\frac{1}{n-1}\sum_{j=1,i\neq j}^{n}\delta_{ij}$.}{\large \par}

{\large{}We also assume that the connections depend entirely on the
two involved vertices and not on other factors. So, $\delta_{ij}'s$
are independent. }{\large \par}

{\large{}The distribution of $d_{i}$can be explicitly obtained by
the results from Wang's Paper$^{[2]}$ }{\large \par}

\textbf{\large{}Proposition: }{\large{}By an easy application of the
Lyapunoff condition regarding to the Central limit theorem for independent
but not identically distributed random variables, we have the large
sample distribution of $\frac{d_{i}}{\sqrt{n}}$ as}{\large \par}

{\large{}$\text{\textsurd}n\frac{\{\frac{1}{n}\sum_{i=1}^{n}(\delta_{ij}\text{\textminus}p_{ij})}{\surd p_{ij}(1\text{\textminus}p_{ij})/n}\underrightarrow{Law}N(0,1).$
as long as $p_{ij}$is bounded away from 0 and 1.}{\large \par}

{\large{}$\text{\textsurd}n\frac{\{\frac{1}{n}\sum_{i=1}^{n}(\delta_{ij}\text{\textminus}p_{ij})}{\surd p_{ij}(1\text{\textminus}p_{ij})/n}=\frac{d'_{i}}{\sqrt{n}}$,
which is a scaled degree-density.}{\large \par}

{\large{}Then}{\large \par}

{\large{}Also note, under this structure}{\large \par}

{\large{}$Cov(d_{i},d_{k})=Cov(\sum_{j=1,i\neq j}^{n}\delta_{ij},\sum_{j=1,k\neq j}^{n}\delta_{kj},)$}{\large \par}

{\large{}$\qquad\qquad=\sum_{j\neq i}\sum_{l\neq k}Cov(\delta_{ij},\delta_{lk})$}{\large \par}

{\large{}Under the condition of independence, we have}{\large \par}

{\large{}$=Cov(\delta_{ik},\delta_{ik})=Var(\delta_{ik}=p_{ik}(1-p_{ik})$}{\large \par}

{\large{}Consider the vector $\underbrace{d`}=(d_{1,}d_{2},.....,d_{n})$}{\large \par}

{\large{}Then the $Cov(\frac{d_{i}}{\sqrt{n}},\frac{d_{j}}{\sqrt{n}})=\frac{p_{ik}(1-p_{ik})}{n}\rightarrow0$
asymptotically.}{\large \par}

{\large{}$Var(d_{i})=Var(\sum_{j=1,i\neq j}^{n}\delta_{ij})$}{\large \par}

{\large{}$\qquad=\sum_{j\neq i}Var(\delta_{ij})$ under independence.}{\large \par}

{\large{}$\qquad=\sum p_{ij}(1-p_{ij})$}{\large \par}

{\large{}\medskip{}
}{\large \par}

{\large{}Then Centrality can be tackled as an eigenvalue $\mathbb{E}=Q\Lambda Q'$.
Limiting arguments are not easily available but see the references
below.}{\large \par}

{\large{}The precision for estimation of $d_{i}$is $\frac{1}{\sum p_{ij}(1-p_{ij})}$
\{I cannot use this to formulate the sampling argument\}}{\large \par}

\subsection{Planarity etc.}

{\large{}One important question: $1-P(planarity)\leq P(K_{5})+P(K_{3,3})$,
need to compute these. So if $P(K_{5})\leq\alpha_{5}$ and $P(K_{3,3})\leq\alpha_{3,3}$
s.t. $\alpha_{5}+\alpha_{3,3}\leq\alpha$ then we have one definition
of \textquotedbl{}planar with $(1-\alpha)$ confidence.\textquotedbl{}}{\large \par}

\subsubsection{{\large{}Limiting Distribution of Adjacency Matrix}}

{\large{}The concept of a limiting distribution of the adjacency matrix,
when n$\rightarrow\infty$ will be very helpful for s. In this context
we recall the following:}{\large \par}

\textbf{\textit{\large{}Bose and Sen (2008)}}{\large{} deal with real
symmetric matrices. If \textgreek{l} is an eigenvalue of multiplicity
m of an n \texttimes{} n matrix $A_{n}$ , then the Empirical Spectral
Measure puts mass m/n at \textgreek{l}. Note that if the entries of
$A_{n}$ are random, then this is a random probability measure. If$\lambda_{1}$,
$\lambda_{2}$, . . . , \textgreek{l}$_{n}$ are all the eigenvalues,
then the empirical spectral distribution function (ESD) F$^{A_{n}}$
of $A_{n}$ is given by
\[
F^{A_{n}}(x)=\frac{1}{n}\sum_{i=i}^{n}I(\lambda_{i}\leq x)
\]
Let \{$A_{n}$ \} be a sequence of square matrices with the corresponding
ESD \{F$^{A_{n}}$\}. The Limiting Spectral Distribution (or measure)
(LSD) of the sequence is defined as the weak limit of the sequence
\{F$^{A_{n}}$\}, if it exists. If \{$A_{n}$\} are random, the limit
is in the \textquotedblleft almost sure\textquotedblright{} or \textquotedblleft in
probability\textquotedblright{} sense.}{\large \par}

{\large{}The relevant example for us is the }\textbf{\textit{\large{}Wigner
matrix}}{\large{}. In its simplest form, a Wigner matrix W$_{n}$
of order n is an n \texttimes{} n symmetric matrix whose entries on
and above the diagonal are i.i.d. random variables with zero mean
and variance one. Denoting those i.i.d. random variables by \{x$_{ij}$:
1 \ensuremath{\le} i \ensuremath{\le} j \}, we can visualize the Wigner
matrix as }{\large \par}

{\Large{}
\[
W_{n}=\left[\begin{array}{cccccc}
x_{11} & x_{12} & x_{13} & \cdots & x_{1(n-1)} & x_{1n}\\
x_{12} & x_{22} & x_{23} & \cdots & x_{2(n-1)} & x_{2n}\\
 &  &  & \vdots\\
x_{1n} & x_{2n} & x_{3n} & \cdots & x_{n(n-1)} & x_{nn}
\end{array}\right]
\]
}{\Large \par}

{\large{}The semi-circular law W arises as the LSD of n$^{\text{\textminus}1/2}$
W$_{n}$. It has the density function 
\[
p_{W}(s)=\begin{cases}
\begin{array}{c}
\frac{1}{2\pi}\sqrt{4-s^{2}}\\
0
\end{array} & \begin{array}{c}
if\:|s|\leq2\\
otherwise
\end{array}\end{cases}
\]
All its odd moments are zero. The even moments are given by 
\[
\beta_{2k}(W)=(2k)!/k!(k+1)!
\]
}{\large \par}

\textbf{\large{}Theorem}{\large{}: Let \{w$_{ij}$: 1 \ensuremath{\le}
i \ensuremath{\le} j, j \ensuremath{\ge} 1\} be a double sequence
of independent random variables with E(w$_{ij}$) = 0 for all i \ensuremath{\le}
j and E(w$_{ij}^{2}$) = 1 which are either (i) uniformly bounded
or (ii) identically distributed. Let W$_{n}$be an n\texttimes n Wigner
matrix with the entries w$_{ij}$. Then with probability one, F$^{n^{\text{\textminus}1/2}W_{n}}$converges
weakly to the semicircular law.}{\large \par}

{\large{}We will use this result...}{\large \par}

\medskip{}

{\large{}Using Bose and Sen, 2008: the eigenvalues of $K_{5}$ are
(4, -1, -1, -1, -1) and that of $K_{3,3}$ are (3, 0, 0, 0, 0, -3).
}\\
{\large{}If we are able to work out the }\emph{\large{}ESD}{\large{}
for $((\frac{\delta_{ij}-p_{ij}}{\sqrt{p_{ij(1-p_{ij})}}}))$ then
if we can probabilistically bound it (i) above by 4, then $K_{5}$
is ruled out; (ii) below by -3 to rule out $K_{3,3}$. Then we have
a test for }\emph{\large{}planarity}{\large{}.}{\large \par}

\section{\emph{\large{}Sampling Strategy}}

{\large{}If we use attribute information for nodes, then one approach
could be as follows: }{\large \par}
\begin{enumerate}
\item {\large{}Use }\emph{\large{}Dirichlet}{\large{} process on attributes
(as in }\textbf{\large{}Sethuraman}{\large{}, }\textbf{\large{}1994}{\large{}).
}\\
{\large{} Prior sampling for computing posterior distribution. }{\large \par}
\item {\large{}Now we can use either }\emph{\large{}Markov Random Fields
model }{\large{}(as in }\textbf{\large{}Jordan and Wairight,2008}{\large{})
or }\emph{\large{}Exponential Random Graph}{\large{} model (as in
}\textbf{\large{}Christakis et. al}{\large{} }\textbf{\large{}2011}{\large{})
for edges: $(\delta_{ij}|X_{i},X_{j})$. }\\
{\large{} Now $\delta_{i}\frac{\wedge}{-}\sum_{j}\delta_{ij}|X_{i},X_{j}$
}\\
{\large{} So we use sample estimate for $d_{i}$, $\hat{d_{i}}$ (}\textbf{\large{}use
Lovasz type strategy}{\large{}). }\\
{\large{} Now use }\emph{\large{}EM}{\large{} to estimate $\delta_{ij}|\hat{d_{i}},X_{i},X_{j}\;\forall j$
(to set this up as in }\textbf{\large{}SSSD}{\large{}, }\textbf{\large{}2014}{\large{})}%
\footnote{{\large{}See }\textbf{\large{}Saad, Basar et. al.}{\large{} for a
different but interesting application of such technique in modelling
sharing of information for more efficient estimation.}%
}{\large{}}\\
{\large{} }{\large \par}
\item {\large{}Now we extrapolate for $|V|\rightarrow\infty$ }\\
{\large{} -- this is where the limit theorems will need to be formulated.
Target is to establish }\emph{\large{}weak laws}{\large{} and }\textit{\large{}Central
Limit Theorems}{\large{}.}\\
{\large{} Assumptions on similarity as $n\rightarrow\infty$. }\\
{\large{} Approximate finite basis with dimension $k(n)$ s.t. $\frac{k(n)}{n}\rightarrow0$
as $n\rightarrow\infty$. In fact target $\frac{k(N)}{N}\rightarrow0$
with sample size $N$. }\\
{\large{} So need to bound two approximations. }{\large \par}
\end{enumerate}

\section*{{\large{}Lovasz type sampling strategy}}
\begin{itemize}
\item {\large{}Start with any vertex, call it 1, observation available is
$\{X_{1}\}$. }{\large \par}
\item {\large{}Crawl all connections (neighbours) of 1, observation available
$\{d_{1},X_{i}$ for all neighbours $\}$ }{\large \par}
\item {\large{}Randomly select some of the neighbours of 1. Crawl all connections
of them ... }{\large \par}
\item {\large{}After these two layers, we will have data on $\{d_{i},X_{i}\}$
for $i$ belonging to sampled vertices of these two layers (say first
N) and $\{X_{j}\}$ for connections $j$ of them (say N+1 to N+M). }{\large \par}
\end{itemize}
{\large{}This can be visualised in terms of the data structure given
below. Here the first N rows \& columns will be completely known.
For the next M rows \& columns, $X_{j}$ will be known and some of
the $d_{ij}$'s will be known (loops back). Now from this data the
analysis will begin.}{\large \par}

\smallskip{}

\subsection{Algorithm}

{\large{}Assuming $X_{n}$to be a discrete valued random variable,
say taking values from the set $S=\{s_{1},s_{2},...,s_{k}\}$. There
are multiple iid copies of vertices with the attribute value $s_{i}$.
Since by our assumption the behavior of the vertex is completely determined
by its attribute value, then the centrality of all vertices with same
attribute value should be same. Thus $p_{ij}$also takes finitely
many values, and it is enough to look at the matrix consisting of
$p_{ij}=P(\delta_{ij}=1|X_{i}=s_{k},X_{j}=s_{l})$ for all $k,l=1,..,n$}{\large \par}

{\large{}We attempt to apply a strategy similar to Lovasz. We randomly
sample a vertex and consider its depth-2 neighborhood. By the proportion
of his connections to different vertices with different attribute
values, we get an estimate of $p_{ij}$. If $\exists m,n$ such that
$p_{mn}is$ not estimated from the sample,but $p_{mr}$and $p_{nr}$are
for some r, we note that $d$ being a metric we have the triangle
inequality $d(s_{m},s_{n})\leq d(s_{m},s_{r})+d(s_{r},s_{n})$}{\large \par}

{\large{}So, we have $p_{mn}\geq p_{mr}+p_{nr}$ which provides a
lower bound for $p_{mn}.$We can have $p_{mn}\geq\underset{r}{sup}\{p_{mr}+p_{nr}\}$.
So an iterative updation may be done of the lower bound. If in our
final sample it is still not estimated we can take $p_{mn}\sim U(\underset{r}{sup}\{p_{mr}+p_{nr}\},1)$ }{\large \par}

\subsection*{A. Sampling Algorithm:}

{\large{}Step 1. Select at random $n$ vertices from all the vertices
in the graph. }{\large \par}

{\large{}Step 2. Consider the proportion of vertices that are from
the $i^{th}$class, $i=1,...,k$. Call it $s_{i}^{0}.$}{\large \par}

{\large{}Step 3. For the connection probability of elements of Class
$i$ and Class $j$, use the first measure:}{\large \par}

\begin{center}
\textit{\LARGE{}$\hat{p_{ij}^{0}}=\frac{\#connection\, present\, among\, elements\, of\, class\, i\, andd\, j}{total\, possible\, connections\, from\, i\, to\, j}=\frac{n_{ij}}{s_{i}^{0}s_{j}^{0}n^{2}}$}
\par\end{center}{\LARGE \par}

{\large{}Step 4: For each of the vertices chosen at Step 1, say the
selected vertex is from class $k$ and the neighbours are from classes
$\alpha_{1k,},....,\alpha_{n_{k}k}.$ Pick one of these neighbours,
say the$j^{th}$one from class $i$ at random with probability $p_{k\alpha_{ik}}$.}{\large \par}

{\large{}Step 5:Using the information from the arbitrarily chosen
neighbours, repeat Step 2 to get $s_{i}^{1}$, and calculate $\widehat{p_{ij}^{1}}$similarly.}{\large \par}

{\large{}Step 6. Calculate $q_{ij}=\beta\hat{p_{ij}^{0}}+(1-\beta)\widehat{p_{ij}^{1}}$
where $\beta\in(0,$1). Report it as the probability.}{\large \par}

{\large{}Step 7: If $\exists m,n$ such that $p_{mn}is$ not estimated
from the sample,but $p_{mr}$and $p_{nr}$are for some r, we note
that $d$ being a metric we have the triangle inequality $d(s_{m},s_{n})\leq d(s_{m},s_{r})+d(s_{r},s_{n})$}{\large \par}

{\large{}So, we have $p_{mn}\geq p_{mr}+p_{nr}$ which provides a
lower bound for $p_{mn}.$We can have $p_{mn}\geq\underset{r}{sup}\{p_{mr}+p_{nr}\}$.
We take $q_{mn}\sim U(\underset{r}{sup}\{p_{mr}+p_{nr}\},1)$ }{\large \par}

\subsection*{B. Generation of the Adjacency Matrix from our Sampling Scheme:}

{\large{}Assume that the graph size is unknown}{\large \par}

{\large{}First assign the vertex $i$ to a class $C_{i}$ by generating
the random variable $X$ from the discrete distribution of the standardized
$s_{i}$.}{\large \par}

{\large{}Then, if vertex $i$ is from $C_{k}$and if vertex $j$ is
from $C_{r},$then $((a_{ij}))\sim Bernoulli(q_{kr})$}{\large \par}

{\large{}Note:}{\large \par}

{\large{}$Prob(a_{ij}=1)=\sum_{k,r}Prob(a_{ij}=1,i\epsilon C_{k},j\epsilon C_{r})=\sum_{k,r}Prob(a_{ij}=1|i\epsilon C_{k},j\epsilon C_{r})Prob(i\epsilon C_{k},j\epsilon C_{r})=\sum_{k,r}Prob(a_{ij}=1|i\epsilon C_{k},j\epsilon C_{r})Prob(i\epsilon C_{k})Prob(j\epsilon C_{r})$}{\large \par}

{\large{}$=\sum_{k,r}q_{kr}s`_{k}s`_{r}$}{\large \par}

\subsection*{C. The resulting data structure \protect \\
}

{\large{}The above sampling scheme will give rise to data which is
a finite subgraph of the original graph in the following structure:}{\large \par}

{\large{}}%
\begin{tabular}{c|cccccc|c}
 & {\large{}$X_{1}$ } & {\large{}$X_{2}$ } & {\large{}$X_{3}$ } & {\large{}\ldots{}} & {\large{}$X_{k}$ } & {\large{}\ldots{}} & {\large{}$\overrightarrow{d}$ }\tabularnewline
 & {\large{}1 } & {\large{}2 } & {\large{}3 } & {\large{}\ldots{}} & {\large{}k } & {\large{}\ldots{}} & \tabularnewline
\hline 
{\large{}1 } & {\large{}0 } & {\large{}$\delta_{12}$ } & {\large{}$\delta_{13}$ } & {\large{}\ldots{}} & {\large{}$\delta_{1k}$ } & {\large{}\ldots{}} & {\large{}$d_{1}$ }\tabularnewline
{\large{}2 } & {\large{}$\delta_{12}$ } & {\large{}0 } & {\large{}$\delta_{23}$ } & {\large{}\ldots{}} & {\large{}$\delta_{2k}$ } & {\large{}\ldots{}} & {\large{}$d_{2}$ }\tabularnewline
{\large{}3 } & {\large{}$\delta_{13}$ } & {\large{}$\delta_{23}$ } & {\large{}0 } & {\large{}\ldots{}} & {\large{}$\delta_{3k}$ } & {\large{}\ldots{}} & {\large{}$d_{3}$ }\tabularnewline
{\large{}$\vdots$ } &  &  & {\large{}$\ddots$ } &  &  &  & {\large{}$\vdots$}\tabularnewline
{\large{}$k$ } &  &  & {\large{}\ldots{}} &  & {\large{}0 } & {\large{}\ldots{}} & {\large{}$d_{k}$ }\tabularnewline
{\large{}$\vdots$ } &  &  &  &  &  &  & {\large{}$\vdots$}\tabularnewline
\hline 
\end{tabular}{\large \par}

\section{Results}
\begin{itemize}
\item {\large{}Simulation Results}{\large \par}
\item {\large{}Comparison with existing methods (degree centrality etc.)}{\large \par}
\item {\large{}Test results for Planarity etc.}{\large \par}
\end{itemize}

\section{Discussion and Conclusions}

\begin{onehalfspace}
\emph{\large{}6.1: On Infinite Graph Spectrums:}{\large \par}

{\large{}In Bose and Sen$^{[1]}$, the questions about the spectral
decomposition of infinite dimensional matrices are tackled, with results
derived for the Winger Matrix. However, the realised Adjacency Matrix
for our model does not have an iid structure of rows, as the value
on the $(i,j)^{th}$ row depends totally on the class of the $i^{th}$element
and the $j^{th}$element. The submatrices of the form given below
are Winger (ie, the rows are generated from an iid process) and for
the individual blocks we can obtain the limiting spectral distribution
(LSD), which is the limit of $F^{A_{n}}(x)=n^{-1}\sum1I(\lambda_{i}\le x).$}{\large \par}

{\large{}$A_{n}=$$\begin{array}{ccccc}
 & c_{j1} & c_{j2} & c_{j3} & ...\\
c_{i1}\\
c_{i2}\\
c_{i3}\\
...
\end{array}$}{\large \par}

{\large{}This, while can give an idea about the large sample centrality
of the classes, seems to fail to generalize to give overall graph
spectra.}{\large \par}

{\large{}Our overall graph has finitely many blocks of the above form,
with each block of infinite size. If we can prove the result for an
adjacency matrix with two classes, then we can extend the result to
finitely many blocks.}{\large \par}

\emph{\large{}6.2: Weighted Graph}{\large \par}

{\large{}If instead of considering the actual adjacency matrix we
consider the weighted graph adjacency matrix, where link weights are
the probabilities of connection between the two vertices, the underlying
weighted graph is connected.}{\large \par}

{\large{}However, with such a notion of weighted graph the degree
of a vertex of an infinite graph $d_{i}=\sum_{j}p_{ij}$ is always
going to be infinite, as it will connect to all other vertices with
some nonzero probability, and is basically a sum of infinitely many
values of $p_{ij}\neq0$ for atleast one $j$ (the graph being infinite)
and hence converges to infinity.}{\large \par}

{\large{}Another issue with such a setup would be that all vertices
of the same class should theoretically have the same centrality, as
connections are dependent only on class properties and not the individual
vertices themselves. Thus, we may end up characterizing an ``influential''
group of people rather than identifying any one individual- which
makes sense from a marketing/ SNA perspective. If our attributes are
fine enough, then the number of classes will be high with low class
size for most, leading to zeroing in on one influential person compared
to the rest.}{\large \par}

{\large{}The notion of $p_{ij}$bounded away from zero arises from
the notion that we may have incomplete information about the attributes,
so we cannot with certainity say who will connect to/avoid whom. }{\large \par}

\emph{\large{}6.3: Spectrum of the Weighted Graph}{\large \par}

{\large{}In such a case, we may again note that any finite column-truncated
(respectively row-truncated) subgraph has repeated rows (respectively
column truncated), and finding the eigenvalues of any finite dimensional
subgraph with repeated rows is equivalent to finding the eigenvalue
of a transformed lower dimensional matrix, as follows.}{\large \par}

{\large{}In general, if I is a set of rows which are identical, then
let $v_{I}$ be the vector which is $1/\sqrt{|I|}$ on the coordinates
in I and 0 elsewhere. The $v_{I}$ are orthonormal, complete them
to an orthonormal basis by adding vectors $w_{j}$. Then A will annihilate
the $w_{j}$ and will take $Span(v_{I})$ to itself. The matrix of
endomorphism of $Span(v_{I})$ will have entries that look like$\sqrt{IJ}aij,$
with$i\in I$ and $j\in J$. So it suffices to compute the eigenvalues
of this finite matrix, and the rest are 0.}{\large \par}

{\large{}For every finite subgraph of the matrix say of dimension
$n\times n$, calculating the subgraph spectra is thus equivalent
to calculating the graph spectra of a smaller transformed matrix,
i.e, if we have k classes, we can simply compute the eigenvalues of
the finite matrix of dimension $\tbinom{k}{2}\times\tbinom{k}{2}.$}{\large \par}
\end{onehalfspace}

\begin{onehalfspace}

\section*{{\large{}\pagebreak{}}}
\end{onehalfspace}

\section*{{\large{}References}}
\begin{enumerate}
\item {\large{}Another look at the moment method for large dimensional random
matrices, Arup Bose (Indian Statistical Institute) and Arnab Sen (University
of California, Berkeley), Electronic Journal of Probability 2008,
http://ejp.ejpecp.org/article/download/501/706 }{\large \par}
\item {\large{}On the Number of Successes in Independent Trials, Y.H. Wang
(Concordia University)}{\large \par}
\item {\large{}C. Hubler, H. P. Kriegel, K. Borgwardt, Z. Ghahramani, Metropolis
algorithms for representative subgraph sampling, in: Proceedings of
the 8th International Conference on Data Mining, IEEE, 2008, pp. 283
- 292. }{\large \par}
\item {\large{}H. Sethu, X. Chu, A new algorithm for extracting a small
representative subgraph from a very large graph, e-print arXiv:1207.4825. }{\large \par}
\item {\large{}N. Blagus, L. Subelj, G. Weiss, M. Bajec, Sampling promotes
community structure in social and information networks, Physica A
432 (2015) 206 - 215. }{\large \par}
\item {\large{}J. Leskovec, C. Faloutsos, Sampling from large graphs, in:
Proceedings of the 12th ACM SIGKDD International Conference on Knowledge
Discovery and Data Mining, ACM, 2006, pp. 631 - 636. }{\large \par}
\item {\large{}S. H. Lee, P. J. Kim, H. Jeong, Statistical properties of
sampled networks, Phys. Rev. E 73 (1) (2006) 016102. }{\large \par}
\item {\large{}N. Blagus, L. Subelj, M. Bajec, Assessing the effectiveness
of real-world network simplification, Physica A 413 (2014) 134 - 146. }{\large \par}
\item {\large{}N. Ahmed, J. Neville, R. R. Kompella, Network sampling via
edge-based node selection with graph induction, Tech. rep., Purdue
University (2011).}{\large \par}
\item {\large{}I. Benjamini, L. Lovász: Global Information from Local Observation,
{[}Proc. 43rd Ann. Symp. on Found. of Comp. Sci. (2002) 701-710.{]}}{\large \par}
\item {\large{}Sourabh Bhattacharya, Diganta Mukherjee, Sutanoy Dasgupta
and Soumendu Sundar Mukherjee, \textquotedblleft A Model for Social
Networks\textquotedblright , mimeo, ISI, December 2014.}{\large \par}
\item {\large{}Christakis,N., J. Fowler ,G. W. Imbens and K. Kalyanaraman
(2011), An Empirical Model for Strategic Network Formation, Working
Paper, National Bureau of Economic Research, Cambridge, MA }{\large \par}
\item {\large{}W Saad, Z Han, M Debbah, A Hjorungnes, T Basar, Coalitional
game theory for communication networks, Signal Processing Magazine,
IEEE 26 (5), 77-97.}{\large \par}
\item {\large{}Koushiki Sarkar, Abhishek Ray and Diganta Mukherjee,\textquotedblleft Impact
of Social Network on Financial Decisions\textquotedbl{}, forthcoming
in Studies in Microeconomics, 2015. }{\large \par}
\item {\large{}J. Sethuraman A constructive definition of Dirichlet priors
Statistica Sinica, 4 (1994), pp. 639\textendash 650}{\large \par}
\item {\large{}M. J. Wainwright and M. I. Jordan (2008). Graphical models,
exponential families, and variational inference. Foundations and Trends
in Machine Learning, Vol. 1, Numbers 1--2, pp. 1--305, December 2008.}\end{enumerate}

\end{document}